\documentclass[pdflatex,sn-mathphys-num]{sn-jnl}

\usepackage{graphicx}%
\usepackage{multirow}%
\usepackage{amsmath,amssymb,amsfonts}%
\usepackage{amsthm}%
\usepackage{mathrsfs}%
\usepackage[title]{appendix}%
\usepackage{xcolor}%
\usepackage{textcomp}%
\usepackage{manyfoot}%
\usepackage{booktabs}%
\usepackage{algorithm}%
\usepackage{algorithmicx}%
\usepackage{algpseudocode}%
\usepackage{listings}%

\usepackage{dsfont}
\usepackage{natbib}
\usepackage{url}
\usepackage{epstopdf}
\usepackage{color}
\usepackage{fullpage}
\usepackage{amsmath,amsthm,amsfonts,amssymb,amscd}
\usepackage{lastpage}
\usepackage{enumerate}
\usepackage[shortlabels]{enumitem}
\usepackage{fancyhdr}
\usepackage{mathrsfs}
\usepackage{xcolor}
\usepackage{graphicx}
\usepackage{listings}
\usepackage{hyperref}
\usepackage{subcaption}

\newcommand{\be}{\begin{equation}}
\newcommand{\ee}{\end{equation}}
\newcommand{\bea}{\begin{eqnarray}}
\newcommand{\eea}{\end{eqnarray}}

\newcommand{\ket}[1]{|#1\rangle}

\newcommand{\pp}{P_{\phi}}
\newcommand{\rp}{\rho_{\phi}}

\newcommand{\ml}{M_{\lambda}}

\begin{document}

\title[Article Title]{Semiclassical polymer field with a cubic potential in cosmology}

\author[1,2]{\fnm{Ahsan} \sur{Mujtaba}}\email{ahsan.mujtaba@nu.edu.kz}

\author*[2]{\fnm{Syed Moeez} \sur{Hassan}}\email{syed\_hassan@lums.edu.pk}

\affil[1]{Physics Department \& Energetic Cosmos Laboratory, Nazarbayev University,\\
Astana 010000, Qazaqstan}

\affil*[2]{Department of Physics, Syed Babar Ali School of Science and Engineering, Lahore University of Management Sciences, Lahore 54792, Pakistan}

\abstract{We study the semiclassical dynamics of a polymer quantized scalar field with a cubic potential in cosmology. The cosmological spacetime is chosen to be homogeneous and isotropic, and we work in the polymer quantization scheme where the field operator does not exist directly. The dynamics are studied in the `e-fold' time gauge, where a choice of time gauge fixing, in terms of the scale factor, is made at the classical level. The cubic potential is interesting to study because it is unbounded from below. We find that polymer quantization regularizes this potential, and leads to periods of slow-roll as well as exactly de-Sitter inflation.}

\keywords{}

\maketitle

\section{Introduction}

Cosmic inflation is often studied as a model of the early universe. It is characterized by a period of accelerated expansion of the universe, and was proposed to solve the `Horizon' and `Flatness' problems (among others) of the standard big bang cosmology \cite{guth1981inflationary, langlois2005inflation}. Numerous models of inflation have been presented in the literature, with expansion sourced by a scalar field, multiple scalar fields, cosmic fluids, and even fields from the standard model. In addition to inflation in the early universe, we also have evidence that the universe has recently entered an accelerated phase of expansion, which might continue into the infinite future \cite{Planck:2018vyg}.

While inflation addresses many problems of the standard big bang model, it does not resolve the singularity present at early times, as it treats gravity classically. It is expected that a quantum theory of gravity might cure gravity of such singularities. Loop Quantum Gravity (LQG) is one such program that attempts to quantize gravity, and Loop Quantum Cosmology (LQC) is its incarnation in cosmological settings. Indeed, LQC models do predict a resolution of this early singularity \cite{Ashtekar:2011ni, Agullo:2016tjh, Agullo:2023rqq}, and imprints of this quantization on the inflationary scenario have also been studied \cite{bojowald2015loop, tsujikawa2004loop}.

Polymer quantization is a method of quantization motivated from LQG techniques, and distinct from standard `Schrodinger' quantization, that features a discreteness built into the theory at a fundamental level \cite{Ashtekar:2002sn, Halvorson_2004}. Polymer quantization has been studied in a quantum mechanical context \cite{Corichi:2007tf, corichi2007hamiltonian}, as well as in a field theory context applied to matter scalar fields \cite{Laddha:2010hp, Hossain:2009vd, Ashtekar:2002vh}. The dynamics of a polymer quantized scalar field coupled to a homogeneous and isotropic spacetime have been analyzed before \cite{hossain2010nonsingular, hassan2015polymer, hassan2017semiclassical, ali2017natural} with the main result that polymer quantization of the matter sector leads to periods of inflationary expansion of the universe.

In this paper we study the semi-classical dynamics arising from a polymer quantized scalar field living on a cosmological background. The background is taken to be homogeneous and isotropic, and semi-classical dynamics are obtained by computing the expectation value of the scalar field Hamiltonian on this background, in a suitably chosen semi-classical Gaussian state. Our model is distinct from the ones presented in \cite{hossain2010nonsingular, hassan2015polymer, hassan2017semiclassical}, in that we use the scheme where the momentum translations are fundamentally discrete (as opposed to the field translations being fundamentally discrete - the two are not equivalent), and is different from the one in \cite{ali2017natural}, in that we choose the scalar field to be massive, with a cubic self interaction. The cubic potential is interesting to study because, unlike other models discussed before, the cubic potential is unbounded from both above \emph{and} below.

In the next section, we present our model and briefly review the method of polymer quantization. Our main results appear in Sec \ref{sec-dyn}, and we summarize in Sec \ref{sec-sum}. We work in $c=\hbar=8\pi G = 1$ units throughout.

\section{\label{sec-model} The model}

We start with a massive scalar field with a cubic potential minimally coupled to a spatially flat, homogeneous and isotropic Friedmann-Lemaitre-Robertson-Walker (FLRW) universe. The action for this theory is given by,
\be
S=V_0 \int dt \left(P_{a}\dot{a}+\pp\dot{\phi}- \mathcal{N} (H_G + H_{\phi}) \right),
\ee
where, $a$ is the scale factor, $P_a$ is its canonically conjugate momentum, $\phi$ is the scalar field, $\pp$ is its conjugate momentum,
\be
H_G = -\dfrac{P_{a}^{2}}{12a}
\ee
is the gravitational Hamiltonian,
\be
H_{\phi} = \dfrac{P_{\phi}^{2}}{2a^{3}}+a^{3}V(\phi)
\ee
is the scalar field Hamiltonian with $V(\phi) = \frac{1}{2} m^2 \phi^2 + g \phi^3$ ($m$ being the scalar field mass, and $g$ the cubic coupling constant), $V_0$ is the volume of a fiducial cell (i.e., $V_0 = \int d^3x)$, and $\mathcal{N}$ is the lapse function which enforces the Hamiltonian constraint,
\be
\label{ham-constr}
H_G + H_{\phi} \approx 0.
\ee

\begin{figure}[h]
\begin{center}
\includegraphics[width=0.8\linewidth]{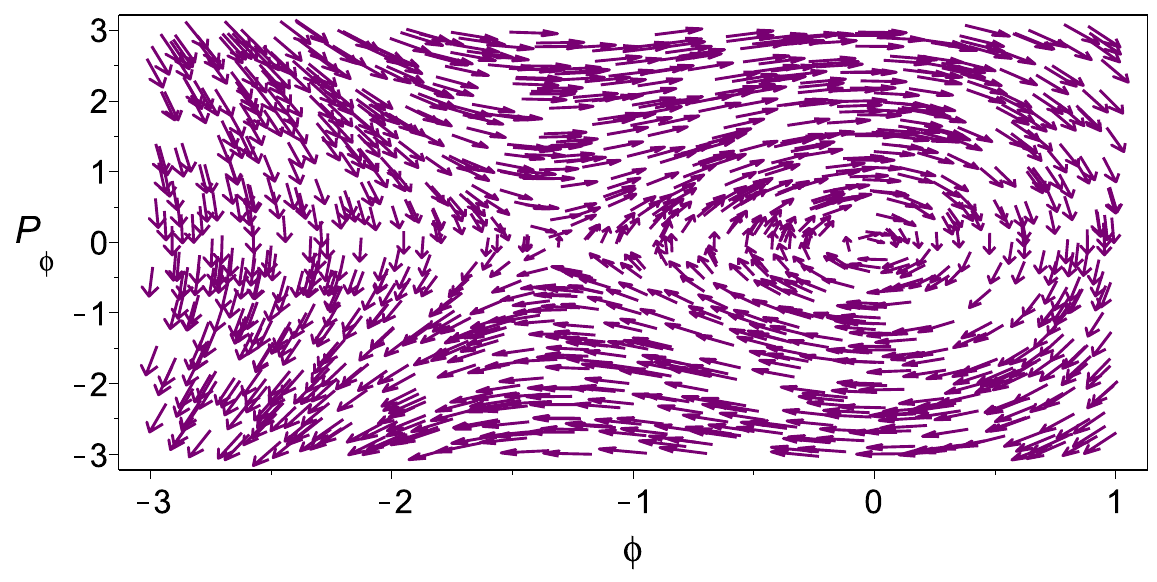}
\end{center}
\caption{\label{fig-phase-classical} Classical phase portrait of a scalar field with a cubic potential. The mass is chosen to be $m=2$, and the cubic coupling $g=1$. Initial conditions are chosen between -3 and 1 for $\phi$, and between -3 and 3 for $\pp (= d\phi/dt)$. Both the stable equilibrium (0, 0), and the unstable equilibrium (-4/3, 0) are clearly visible.}
\end{figure}

To provide a quick overview of the cubic potential, we note that $V(\phi)$ is unbounded from both above \emph{and} below, and features local minima and maxima at $\phi=0$, and $\phi=-m^2/3g$ respectively. A classical phase portrait for this potential is shown in Fig \ref{fig-phase-classical}.

The FLRW line element is invariant under spatial dilations $x \rightarrow lx, a \rightarrow a/l$ where $l$ is some arbitrary length scale. This allows us to define scale invariant physical variables that do not depend on the choice of the arbitrary $V_0$ as,
\be
a = V_0^{1/3}a,~~ P_a = V_0^{2/3}P_a,~~ \phi = \phi,~~ \pp = V_0 \pp.
\ee
In what follows, we work with these scale invariant variables.

We look at the dynamics of this system in the `e-fold' time gauge, where we fix a time gauge at the classical level, and solve the Hamiltonian constraint to get a non-vanishing physical Hamiltonian. We choose the number of e-folds as time,
\be
t = N = \ln{a}.
\ee
The momentum conjugate to this time variable is $aP_a$, and solving the Hamiltonian constraint for this conjugate momentum gives us the physical Hamiltonian as,
\be
\label{phys-ham}
H_p = -aP_a = 6 e^{3N} \sqrt{\dfrac{\rp}{3}},
\ee
where,
\be
\label{eqn-rho}
\rp = \dfrac{H_{\phi}}{a^3} = \dfrac{\pp^2}{2a^6} + V(\phi) = \dfrac{\pp^2}{2 e^{6N}} + V(\phi)
\ee
is the scalar field energy density. Our goal is to polymer quantize the scalar field, compute the expectation value of $\rp$ in a suitably chosen semi-classical state, and study the resultant dynamics arising from the physical Hamiltonian above.

\subsection{Polymer quantization}

We now briefly review the method of polymer quantization for completeness, and for details, we refer the reader to \cite{ali2017natural, hassan2015polymer}. The basic variables are taken to be $U_{\lambda} \equiv e^{i\lambda\phi}$ and $\pp$ which satisfy the Poisson bracket relation,
\begin{equation}
    \{U_{\lambda},\pp\}=i\lambda U_{\lambda},
\end{equation}
where $\lambda$ is a free parameter (with dimensions of inverse mass)  and represents the polymer scale, $\ml \equiv 1/\lambda$. We work in the `momentum' basis where the basis states are represented by $|P\rangle$ with the inner product,
\begin{equation}
    \langle P|P'\rangle=\delta_{PP'}.
    \label{eq4}
\end{equation}
where $\delta_{PP'}$ is the generalization of the Kronecker delta function to the real numbers.
The action of the operators, $\widehat{\pp}$ and $\widehat{U_{\lambda}}$, on the basis $\ket{P}$ is given as,
\begin{subequations}
\begin{equation}
    \widehat{\pp}\,|P\rangle=P|P\rangle,
    \label{eq6a}
\end{equation}
\begin{equation}
    \widehat{U}_{\lambda}\,|P\rangle=|P+\lambda\rangle
    \label{eq6b}
\end{equation}
\end{subequations}  
(so $|P\rangle$ is a momentum eigenstate with eigenvalue $P$, and $\widehat{U_{\lambda}}$ acts as a translation operator). The translation operator is not weakly continuous in $\lambda$, meaning that the $\lambda \rightarrow 0$ limit does not exist, and hence, the field operator $\widehat{\phi}$ cannot be defined directly. Instead, the classical expansion of $U_{\lambda}$ motivates an indirect definition,
\begin{equation}
    \widehat{\phi} \equiv \dfrac{1}{2 i \lambda}(\widehat{U}_{\lambda}-\widehat{U}^{\dagger}_{\lambda}).
    \label{redefined scalar field operator}
\end{equation}

With these operator definitions, the scalar field energy density operator is formally given by (after converting the classical expression (\ref{eqn-rho}) into an operator),
\be
\widehat{\rp} = \dfrac{\widehat{\pp}^2}{2 e^{6N}} + \dfrac{1}{2} m^2 \widehat{\phi}^2 + g \widehat{\phi}^3.
\label{rho phi formula}
\ee

We now compute the expectation value of this energy density in a semi-classical Gaussian state given by,
\begin{eqnarray}
    |\psi\rangle &=& \dfrac{1}{\mathfrak{N}}\sum^{\infty}_{k=-\infty}C_{k}\ket{P_k}, \\
    C_{k} &=& \exp\Big[-{\dfrac{(P_{k}-\pp)^{2}}{2\sigma^{2}}}\Big]\exp{(-i P_{k}\phi)},
    \label{gaussian}
\end{eqnarray}
where $\sigma$ is the width of the state, $(\phi,\pp)$ represent the peaking values, and $\mathfrak{N}$ is a normalization constant. The sum is approximated by an integral, and the final result is (again, for details, we refer to \cite{ali2017natural, hassan2015polymer}),
\begin{equation}
    4 \langle \widehat{\rho}_{\phi}\rangle = e^{-6N} \left( 2 \pp^{2} + \dfrac{1}{M_\lambda^2 \Sigma^2} \right) + m^{2} M_{\lambda}^{2}\left(1-e^{-\Sigma^{2}} \cos{2\Theta}\right) + g M_\lambda^3 \left(e^{-9\Sigma^2/4} \cos {3\Theta} - 3 e^{-\Sigma^2/4} \sin{\Theta} \right),
\label{sc-rho}
\end{equation}
where $\Theta \equiv \phi/M_\lambda$ is a variable containing the dynamical scalar field $\phi$, and $\Sigma \equiv (\sigma M_\lambda)^{-1}$ is a constant encoding the effects of the Gaussian state width $\sigma$, and the polymer scale $M_{\lambda}$.

\begin{figure}[h]
\begin{center}
\includegraphics[scale=0.8]{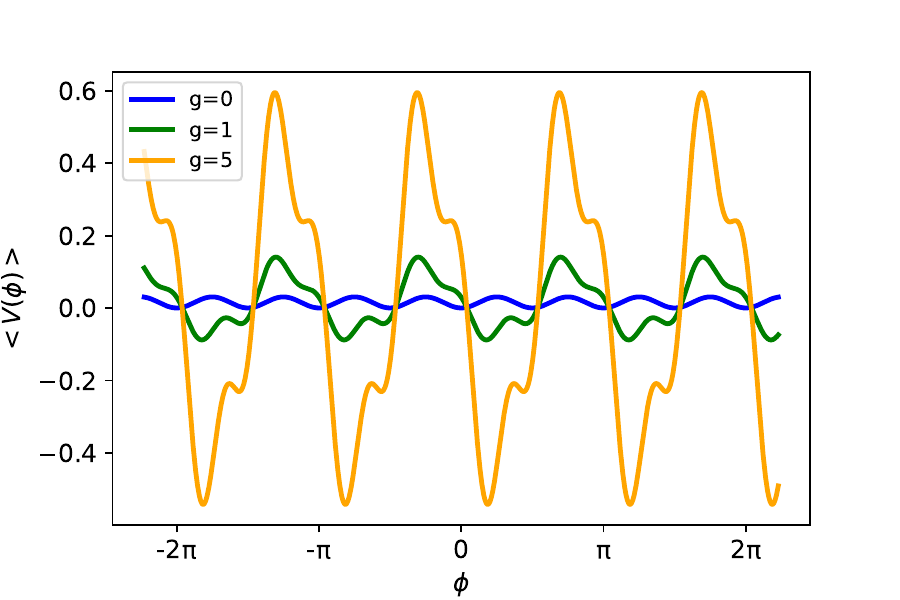}
\end{center}
\caption{\label{fig-effective-pot} The polymer quantized effective potential for various values of the cubic coupling constant $g$ (the $g=0$ case corresponds to the massive scalar field). The scalar field mass is $m=0.5$, polymer scale $M_\lambda = 0.5$, and the semi-classical state width $\sigma = 25$.}
\end{figure}

The effective potential that arises from this scheme is shown in Fig \ref{fig-effective-pot} for various values of the cubic coupling constant $g$. We note some interesting features of this potential: Firstly, the potential is bounded due to polymer quantization from both above and below. Hence polymer quantization serves as a regulator on both ends of the potential (the bottomless coulomb potential has also been regularized using polymer quantization techniques \cite{husain2007quantum}). Secondly, it is periodic, as expected, for all values of $g$. Thirdly, for $g=0$, the potential reduces to the usual (polymer quantized) quadratic potential \cite{ali2017natural}, with repeating minima and maxima. And lastly, for larger values of $g$, we note that there are multiple maxima and minima that occur within a single period. This means that, depending on the initial conditions, there are multiple equilibrium solutions. We now turn to describing the dynamics of this system.

\section{\label{sec-dyn} Semiclassical dynamics of the cubic field}

The semiclassical equations of motion are obtained from the physical Hamiltonian (\ref{phys-ham}), with the classical scalar field energy density replaced with its semi-classical counterpart given in (\ref{sc-rho}),

\begin{equation}
 \dfrac{d {\phi}}{d N}=e^{-3N} \dfrac{ {\pp}}{ {H}},
 \label{dphi/dN} 
 \end{equation}
 \begin{equation}
     \dfrac{d  {\pp}}{d N} = - e^{3 N} \dfrac{{M}_{\lambda}^2}{ {4}} \dfrac{1}{H} \left[2 {m}^2   e^{-\Sigma^2} \sin 2 \Theta - 3 g\left( e^{-9 \Sigma^2/4} \sin 3 \Theta - e^{-\Sigma^2/4} \cos \Theta \right)\right],
     \label{dp/dN}
 \end{equation}

where $H$ is the Hubble parameter, defined as $H = \dot{a}/\mathcal{N}a$, and can be expressed in terms of the scalar field variables by using the Hamiltonian constraint (\ref{ham-constr}), 
 \begin{equation}
 12 {H}^2 = e^{-6N}\left(2 \pp^2 + \dfrac{1}{M_{\lambda}^2 \Sigma^2}\right) + m^{2} M_{\lambda}^{2} \left(1-e^{-\Sigma^{2}} \cos{2\Theta}\right)\\+ g M_\lambda^3\left(e^{-9\Sigma^2/4} \cos 3\Theta -3 e^{-\Sigma^2/4} \sin \Theta \right).
\label{H^2 equation}
\end{equation}
Another variable of interest for studying inflation is the Hubble slow-roll parameter, $\epsilon_H$, defined as,
\begin{equation}
\epsilon_H = -\dfrac{d}{dN} \text{ln} H.
\end{equation}
Inflation happens whenever $\epsilon_H < 1$, and purely de-Sitter inflation occurs when $\epsilon_H = 0$.

Let us look at two interesting limits of these equations: (i) The early time limit ($N \rightarrow -\infty$). In this limit, the Hubble parameter behaves as $H \rightarrow e^{-3N}$, the field momentum becomes constant, the scalar field equation becomes $\dot{\phi} \rightarrow 0$ implying that the field evolves linearly, and the slow roll parameter $\epsilon_H \rightarrow 3$, independent of any parameter values or initial conditions, in agreement with the results of \cite{ali2017natural}. This also implies that the big bang singularity is not resolved, and is pushed into the infinite past. (ii) The late time limit ($N \rightarrow \infty$). The equations of motion themselves do not reduce to a parameter independent limit, however, the Hubble parameter becomes independent of $N$, implying that $\epsilon_H \rightarrow 0$, which indicates an exact de-Sitter inflationary phase.

\begin{figure}[h]
\begin{subfigure}{.5\textwidth}
\begin{center}
\includegraphics[width=1\linewidth]{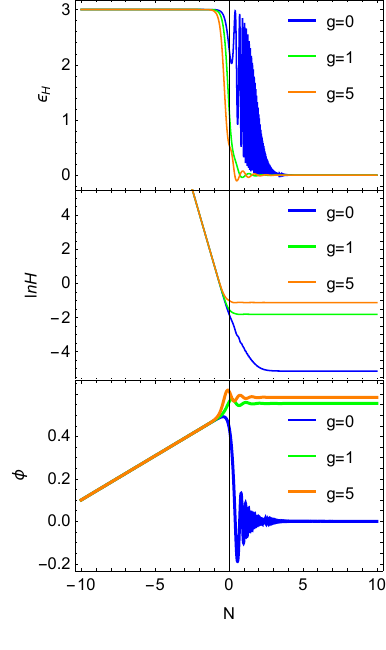}
\end{center}
\end{subfigure}%
\begin{subfigure}{.5\textwidth}
\begin{center}
\includegraphics[width=1\linewidth]{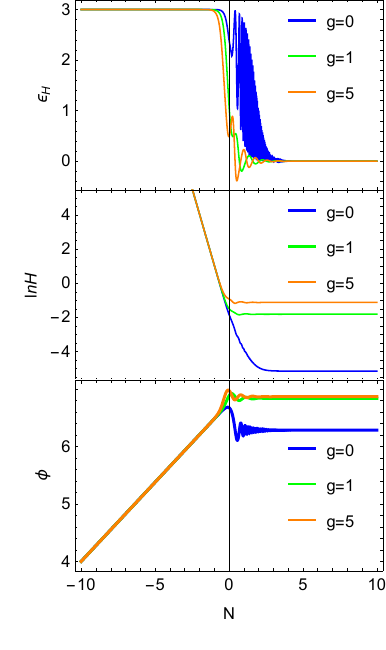}
\end{center}
\end{subfigure}
\caption{\label{fig-phi} Solutions of the semiclassical equations of motions showing the Hubble slow-roll parameter $\epsilon_H$ (\textit{top}), (logarithm of) the Hubble parameter $H$ (\textit{middle}), and the scalar field $\phi$ (\textit{bottom}) against the time $N$, for various values of the cubic coupling constant $g$. The initial conditions were chosen to be, $(\phi, P_{\phi})|_{\text{init}} = (0.1, 0.3)$ \textit{(left panel)}, and $(\phi, P_{\phi})|_{\text{init}} = (4, 2)$ \textit{(right panel)}. The scalar field mass is $m=0.5$, polymer scale $M_\lambda = 0.5$, and the semi-classical state width $\sigma = 25$.}
\end{figure}

We now solve these equations numerically for various choices of initial conditions and parameter values. A typical solution - for different choices of $g$ - is shown in Figure \ref{fig-phi}, that shows the time evolution of the scalar field, (logarithm of) the Hubble parameter, and the slow-roll parameter, for two different choices of initial conditions. We note some salient features of these results: Firstly, the scalar field starts off at some arbitrary position, and given an arbitrary momentum, tends towards the nearest minima. As clear from the effective potential in Fig \ref{fig-effective-pot}, there are multiple values of $\phi$ where this can happen. One major difference between the quadratic ($g=0$), and cubic ($g \neq 0$) cases is that for the quadratic case, the minima are equally spaced, whereas for the cubic case, the minima come in pairs of two (or three if the unstable minimum is counted), and the scalar field can settle in either one of those. Secondly, once the scalar field reaches a minima, it begins oscillating about it, and then the oscillations slowly die out. This can provide a mechanism for re-heating (at different values of the scalar field, depending on the initial conditions). Thirdly, once the scalar field settles in a minima, inflation begins that continues into the indefinite future. This can be seen clearly by looking at the Hubble slow-roll parameter $\epsilon_H$ that reaches zero during this phase, and stays there indefinitely, indicating exact de-Sitter inflation. And finally, before reaching this de-Sitter phase, there is a period of slow-roll inflation characterized by $\epsilon_H < 1$, the duration of which depends on both the initial conditions, and the cubic coupling constant $g$. In  addition, in the cubic case, there are small intervals of $\epsilon_H < 0$ inflation.

\begin{figure}[h]
\begin{subfigure}{.5\textwidth}
\begin{center}
\includegraphics[width=1\linewidth]{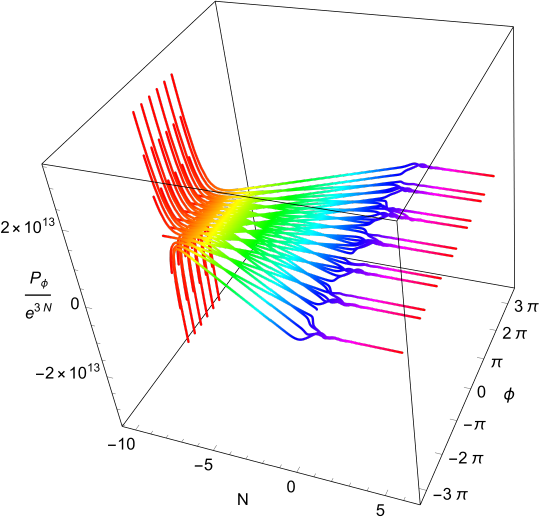}
\end{center}
\end{subfigure}%
\begin{subfigure}{.5\textwidth}
\begin{center}
\includegraphics[width=0.9\linewidth]{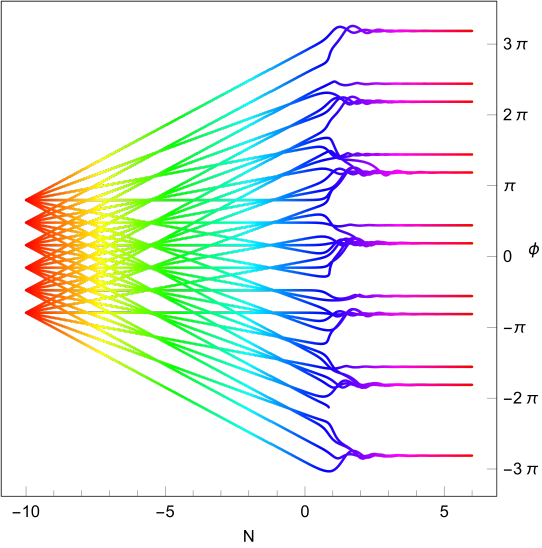}
\end{center}
\end{subfigure}
\caption{\label{fig-phase-3d} \textit{Left:} A three-dimensional phase portrait for the set of equations (\ref{dphi/dN}, \ref{dp/dN}). \textit{Right:} Projection of the three-dimensional phase portrait onto the $(N, \phi)$ plane. The scalar field mass is $m=0.5$, the cubic coupling constant $g=4$, polymer scale $M_\lambda = 0.5$, and the semi-classical state width $\sigma = 15$.}
\end{figure}

To elaborate the role that the choice of initial conditions play, we construct a phase portrait for this system (with $m, g \neq 0$). However, since the physical Hamiltonian (\ref{phys-ham}) (and therefore the equations of motion) explicitly depend on the time $N$, the system is not autonomous, and hence a two-dimensional phase portrait in the $(\phi, \pp)$ plane will in general include crossings of the trajectories. We therefore draw a three-dimensional phase portrait, including $N$ as one of the axes, shown in Fig \ref{fig-phase-3d} (left panel). Time runs from -10 to 6 (with initial conditions provided at $N=-10$), and the initial conditions for $(\phi, \pp)$ are chosen on an equally spaced grid between -2.5 and 3. On the $z$-axis, we display $\pp/e^{3N}$ instead of $\pp$ for ease of visualization. A two-dimensional projection of this onto the $(N, \phi)$ plane is shown in the right panel. We see that for different initial conditions, the scalar field rolls up (or down) the potential, and then settles into a minima (after some oscillations), after which it stays there indefinitely. Depending on the initial conditions, different starting points can lead to the field ending in the same minimum. We also note that, unlike the quadratic case, in the cubic case, the minima occur in pairs of two, a fact that can be traced back to the effective potential (Fig \ref{fig-effective-pot}).

\section{\label{sec-sum} Summary}

We looked at the semi-classical dynamics of a polymer quantized scalar field in a homogeneous and isotropic FLRW spacetime. The scalar field was chosen to be massive, with a cubic potential. Classically, the cubic potential is unbounded from both above and below, and therefore the theory is unstable. Standard quantization of this theory gives a Hamiltonian with no ground state. However, we found that if the field is polymer quantized, the potential becomes bounded from both above and below, thereby providing a stable ground state for the theory.

The scalar field was polymer quantized using a representation in which the field operator $\hat{\phi}$ is not well-defined (and hence the momentum states form a discrete set). This representation is inequivalent to the one in which the momentum operator is not well-defined, and therefore produces qualitatively different results. Semiclassical dynamics were obtained by computing the expectation value of the polymer-quantized scalar field Hamiltonian in a suitably chosen Gaussian state. This effective Hamiltonian is then coupled to the FLRW universe, and the dynamics are obtained from the resulting equations of motion. This approach captures the main features of polymer quantization (at the semi-classical level), without going into the full (and much more complicated) polymer quantum theory.

The polymer quantized cubic scalar field was then coupled to an FLRW universe to study what effects it has on the large-scale cosmological dynamics. This spacetime provides a simple ground for testing many theories, including in particular, inflation. We found that the polymer field produces slow-roll inflation, as well as a late time de-Sitter inflationary phase that continues into the infinite future. The magnitude of this late time inflation (i.e., the Hubble parameter) is controlled by the cubic coupling constant (keeping all other constants fixed). Furthermore, the scalar field also settles into a minimum at late times, and since the (effective) potential is periodic, the location of this minimum depends on the initial conditions. In contrast to the quadratic case however, where there is only one minima in a period, the cubic (effective) potential features minima arising in pairs of two in each period. The initial big bang singularity still persists at earlier times (although it is pushed to the infinite past in the e-fold time gauge) as gravity was treated classically in this model.

Overall, this study elucidates the effects of a polymer quantized scalar field living in a cosmological spacetime, and what differences arise from choosing a non-trivial potential like the cubic one which is unbounded from below. It would be interesting to note what features are introduced if we consider even more exotic potentials, or fields. We leave these and other such questions for future studies.

\backmatter


\bibliography{cubicpoly}

\end{document}